\documentclass[12pt]{article}
\usepackage{graphicx} 
\usepackage[margin=0.9in]{geometry}
\usepackage{amsmath}
\usepackage{bbm}
\usepackage{multirow}
\usepackage{float}
\usepackage{hyperref}
\usepackage[ruled]{algorithm2e}
\usepackage{setspace}
\usepackage[dvipsnames]{xcolor}
\usepackage{subcaption}
\usepackage{authblk}

\title{Sequential importance sampling for thinned count autoregressions via latent Gaussian transformations}
\author[1]{Joey Fingold\thanks{Corresponding author: Joey Fingold, Department of Mathematics and Statistics, University of Guelph, Guelph, Ontario, N1G 2W1, Canada (jfingold@uoguelph.ca)}}
\author[1]{Justin J. Slater}

\affil[1]{Department of Mathematics and Statistics, University of Guelph, Guelph, Ontario, N1G 2W1, Canada}

\date{}

\usepackage[square,authoryear]{natbib}
\allowdisplaybreaks

\begin{document}

\maketitle

\begin{abstract}
    Thinned count autoregressions are popular for modelling infectious disease surveillance data due to their flexibility and interpretability. However, such a model is challenging to fit since it involves high-dimensional and serially correlated integer-valued unknowns. One solution is to consider an analogous continuous-valued surrogate model whose values are post-hoc mapped to integers. This procedure produces biased estimates as inference is performed using samples from such a surrogate model and not the thinned count autoregression itself. In this work, we propose a sequential importance sampling procedure to correct this misspecified model. We demonstrate its validity in a simulation study and its applicability for epidemic curve reconstruction using rotavirus data from Germany and meningococcus data from France. 
\end{abstract}

\textbf{Keywords:} Bayesian inference, count time series, infectious disease modelling, latent Gaussian transformations, sequential Monte Carlo, underreporting

\newpage

\section{Introduction}\label{sec:introduction}

\bigskip
\bigskip

There has been recent interest in using count time series models for infectious disease surveillance data due to their interpretability, flexibility, and their ability to be fit with either maximum likelihood or Bayesian paradigms \citep{Wakefield2019}. An example of such a model, first applied to surveillance data by \citet{held2005}, is the Poisson autoregression (PAR) \citep{Fokianos2009}:
\begin{equation}\label{PAR}
    \begin{aligned}
        X_t|X_{t-1} &\sim \text{Pois}(\lambda_t) \\
        \lambda_t &= \nu + \phi X_{t-1}.
    \end{aligned}
\end{equation}
In this model, case counts, $X_t$, are the number of infected individuals at time $t$. The endemic component, $\nu$, describes cases not attributable to previous cases, while the epidemic component, $\phi X_{t-1}$, describes the persistence of the disease, where $\phi$ can be interpreted as a reproductive rate, as, on average, each case during time step $t-1$ contributes $\phi$ cases to $X_t$. Both $\nu$ and $\phi$ are assumed to be non-negative parameters to be estimated from the data. Other count distributions may be used in place of the Poisson in \eqref{PAR} to allow for conditional overdispersion, such as the negative binomial \citep{zhu2011}. In certain contexts, it may be beneficial to subset these cases into groups based on, for example, geographical region or age \citep{held2005}, thereby extending the model to the multivariate setting, giving the Poisson network autoregression (PNAR) \citep{armillotta2024}:
\begin{equation}\label{PNAR}
    \begin{aligned}
        X_{i,t}|X_{<t} &\sim \text{Pois}(\lambda_{i,t}) \\
        \lambda_{i,t} &= \nu + \phi \sum_{j=1}^m w_{ji} X_{j,t-1},
    \end{aligned}
\end{equation}
where the $w_{ji}$ are constants representing the influence of group $j$ on group $i$, such as proximity \citep{Meyer2014}, social contacts \citep{Meyer2017}, or mobility \citep{slatermobility}. As discussed by \citet{held2012}, among other causes, environmental effects, school holidays, and media attention create seasonality in infectious disease counts. To account for this, one may model $\nu$ and $\phi$ using spatiotemporal processes, allowing both regional components and seasonality, and may be incorporated by letting
\begin{equation}\label{seasonality}
    \begin{aligned}
        \log(\nu_{i,t}) &= \alpha^{(\nu)} + b_i^{(\nu)} + \beta t + \gamma_1^{(\nu)}\sin(\omega t) + \gamma_2^{(\nu)}\cos(\omega t) \\
        \log(\phi_{i,t}) &= \alpha^{(\phi)} + b_i^{(\phi)} + \gamma_1^{(\phi)}\sin(\omega t) + \gamma_2^{(\phi)}\cos(\omega t),
    \end{aligned}
\end{equation}
where the $\alpha$ terms are component intercepts, the $b_i$'s are component and group random effects, and the $\beta$ is a linear trend coefficient. However, such a model is likely misspecified since infectious disease surveillance data is subject to underreporting, as not every infected individual seeks medical care \citep{bracher2021}. To account for this, underreporting can be modelled by considering the observed number of infected individuals at time $t$ in group $i$, $Y_{i,t}$, as a binomially thinned (BT) version of $X_{i,t}$:
\begin{equation}\label{BT}
    Y_{i,t}| X_{i,t} \sim \text{Bin}(X_{i,t},\pi),
\end{equation}
where each case is assumed to be reported with probability $\pi$. This modelling framework also appears in ecology literature as the $N$-Mixture model for measuring abundances. For example, as in \citet{royle2004}, count data corresponding to the number of American redstarts are underreported due to the secretive habits of the bird. 

The challenge is that this model formulation implies that the true counts of infected individuals are serially correlated, integer-valued unknowns. This makes sampling from the posterior defined by this model particularly difficult, as state-of-the-art MCMC methods, like Hamiltonian Monte Carlo (HMC) \citep{Neal2011}, rely on gradient-based optimization and therefore do not apply to the discrete-valued $\{X_{i,t}\}$. Classical methods for hidden Markov models, including the forward-backward algorithm \citep{rabiner1989}, are limited by scalability. Their associated computational burden grows rapidly alongside the dimension of the latent state space, becoming challenging for complex spatiotemporal models. It is also unclear how these methods can be extended to more complicated models, such as those with autoregressions on $\lambda_t$ itself, as seen in \citet{bracher2022}. \citet{bracher2021} provide an approximate maximum likelihood approach to fit such a thinned autoregressive model by considering a model with the same marginal moments yet no underreporting, thereby allowing the back-calculation of parameter estimates. However, their approach assumes that the reporting probability is a known constant. If this $\pi$ is assumed incorrectly, estimates of $\nu$ and $\phi$ will be biased \citep{slater2025}. \citet{slater2025} navigate these issues by proposing a surrogate model defined over latent Gaussian time series $\{Z_{i,t}\}$ which has the same conditional moments, but a continuous parameter space instead, thereby allowing HMC:
\begin{equation}\label{LGPNAR}
    \begin{aligned}
        Y_{i,t}|Z_{i,t} &\sim \text{N}(\pi Z_{i,t}, \pi(1-\pi)Z_{i,t}) \\
        Z_{i,t}|Z_{<t} &\sim \text{N}_{(0,\infty)}(\tilde\lambda_{i,t},\tilde\lambda_{i,t}) \\
        \tilde\lambda_{i,t} &= \nu_{i,t} + \phi_{i,t} \sum_{j=1}^m w_{ji} Z_{j,t-1}.
    \end{aligned}
\end{equation}
These latent Gaussian time series $\{Z_{i,t}\}$ are post-hoc mapped to $\{X_{i,t}\}$ with Poisson distributions parametrized by $\tilde\lambda_{i,t}$ in a transformation inspired by \citet{Jia2023}:
\begin{equation}\label{LGMap}
    X_{i,t} = F_{\tilde\lambda_{i,t}}^{-1}\Bigg [\Phi_{(-\sqrt{\tilde\lambda_{i,t}},\infty)}\Bigg(\frac{Z_{i,t} - \tilde\lambda_{i,t}}{\sqrt{\tilde\lambda_{i,t}}}\Bigg) \Bigg],
\end{equation}
where $F_{\tilde\lambda_{i,t}}^{-1}(u) = \inf \{n : F_{\tilde\lambda_{i,t}}(n) \geq u\}$ for $u \in (0,1)$ is the generalized inverse cumulative distribution function of $X_{i,t}$ and $\Phi_{(-\sqrt{\tilde\lambda_{i,t}},\infty)}(\cdot)$ is the cumulative distribution function of a standard normal random variable truncated below by $-\sqrt{\tilde\lambda_{i,t}}$. The subscript ``$(0,\infty)$" in \eqref{LGPNAR} indicates that the Gaussian distribution is truncated below at zero. This is a requirement of the variance term of $Y_{i,t}|Z_{i,t}$ when fitting the model, as proposing values of $Z_{i,t}$ less than zero will give a negative variance in the density of $Y_{i,t}|Z_{i,t}$. The transformation in \eqref{LGMap} holds since the term inside the square brackets is a uniform(0,1) random variable, and the inverse transform method will recover a random variable with the same distribution as specified by $F(\cdot)$. As the means of these Gaussian approximations, $\pi Z_{i,t}$ and $\tilde\lambda_{i,t}$, get larger, the approximations to the Poisson and Binomial components get better, and as \citet{slater2025} demonstrate, the surrogate posterior becomes a better approximation to the posterior of the target thinned count autoregression.

Although this formulation yields Poisson random variables $\{X_{i,t}\}$, the posterior samples from \eqref{LGPNAR} were not drawn from the desired posterior formulation using \eqref{PNAR} and \eqref{BT}. Consequently, estimates and inference performed using samples from this model will be biased by this model misspecification. In this paper, we propose and demonstrate a novel sequential importance sampling method through which the accuracy of expectations calculated using samples from this surrogate modelling framework is improved. 

The rest of this paper is organized as follows: \autoref{sec:importancesampling} details our proposed sequential importance sampling algorithm and how it corrects the misspecifications of the surrogate model of \citet{slater2025}, \autoref{sec:simstudy} illustrates that our framework is valid through a simulation study, \autoref{sec:casestudies} provides two real-data case studies, demonstrating the applicability of the method, and \autoref{sec:discussion} concludes with limitations and next steps.

\section{Methodology}\label{sec:importancesampling}

\subsection{Importance Sampling}

Suppose $p(\cdot)$ is some target posterior distribution from which it is hard to sample, and $\tilde p(\cdot)$ is a proposal posterior distribution similar to $p(\cdot)$ from which it is easy to sample. We may compute expectations under $p(\cdot)$ as expectations under $\tilde p(\cdot)$ by noting the following relationship:
\begin{align}\label{IS_E}
    \mathbbm{E}_p[h(X,\theta)] &= \mathbbm{E}_{\tilde p}[h(X,\theta)w(X,\theta)],
\end{align}
where $w(X,\theta) = \frac{p(X,\theta|Y)}{\tilde p(X,\theta|Y)}$ are considered weights of the parameters $X$ and $\theta$. This approach is known as importance sampling \citep{kloek_1978}, and motivates a Monte Carlo approximation to \eqref{IS_E} of the form
\begin{equation}\label{SNIS}
    \mathbbm{E}_{\tilde p}[h(X,\theta)w(X,\theta)] \approx \sum_{n=1}^N h(X^{(n)},\theta^{(n)})\tilde w(X^{(n)},\theta^{(n)}),
\end{equation}
where $X^{(n)}$ and $\theta^{(n)}$ are the $n^{\text{th}}$ samples of $X$ and $\theta$ from $\tilde p(\cdot)$ respectively, and $\tilde w(X^{(n)},\theta^{(n)}) = \frac{w(X^{(n)},\theta^{(n)})}{\sum_{k=1}^N w(X^{(k)},\theta^{(k)})}$ is the normalized weight of the $n^{\text{th}}$ sample. This normalization is particularly necessary when $p(\cdot)$ and $\tilde p(\cdot)$ are only known up to proportionality constants, as the integration constants cancel out \citep{givens_hoeting}.
The effectiveness of importance sampling clearly depends on the choice of $\tilde p(\cdot)$. We require (i) that the support of $\tilde p(\cdot)$ includes the entire support of $p(\cdot)$, and (ii) that $p(\cdot)/\tilde p(\cdot)$, for some small number of samples, is not significantly larger than most samples \citep{givens_hoeting}. These two conditions motivate selecting a proposal distribution that is similar to the target distribution. 

Our proposed modelling framework for conducting inference under a thinned count autoregression, defined by \eqref{PNAR} and \eqref{BT} and with posterior $p(\cdot)$, uses this concept of importance sampling by treating the analogous normal-normal surrogate model, as in \eqref{LGPNAR} and \eqref{LGMap}, as a proposal posterior distribution $\tilde p(\cdot)$. This allows us to fit a model with a continuous sample space using HMC, yet still compute estimated expectations under the posterior with a discrete sample space by assigning each sample an importance weight of $\tilde w(X^{(n)},\theta^{(n)})$. Such a procedure corrects model misspecifications of the surrogate model by smoothing expectations through conditioning on the full set of observed data $\{Y_{i,t}\}$. 

Let $\theta = (\nu, \phi, \pi)$ and assume the same priors under both models, $p(\theta) = \tilde p(\theta)$. Under our modelling framework, the importance weights take the form
\begin{align*}
    w(X^{(n)},\theta^{(n)}) \propto \prod_{i=1}^m \Big \{\Big [\prod_{t=2}^T \frac{p(Y_{i,t}|X_{i,t}^{(n)})p(X_{i,t}^{(n)}|X_{<t}^{(n)})}{\tilde p(Y_{i,t}|X_{i,t}^{(n)})\tilde p(X_{i,t}^{(n)}|X_{<t}^{(n)})} \Big] \frac{p(Y_{i,1}|X_{i,1}^{(n)})p(X_{i,1}^{(n)}|\theta^{(n)})}{\tilde p(Y_{i,1}|X_{i,1}^{(n)}) \tilde p(X_{i,1}^{(n)}|\theta^{(n)})} \Big \},
\end{align*}
with an equality replacing the ``$\propto$" under normalization. However, as the number of time steps $T$ and the number of groups $m$ increase, these weights become subject to weight degeneracy. At each time step, less-favoured samples are assigned low weights and preferred samples are assigned high weights, making the number of samples contributing to the expectation progressively smaller. To alleviate this, we define a recursive factorization by indexing each weight by a time step $t$, with $w(X^{(n)},\theta^{(n)}) = w_T^{(n)}$ of the form
\begin{align*}
    w_T^{(n)} &\propto \Big \{\prod_{i=1}^m \Big [\prod_{t=2}^T \frac{p(Y_{i,t}|X_{i,t}^{(n)})p(X_{i,t}^{(n)}|X_{<t}^{(n)})}{\tilde p(Y_{i,t}|X_{i,t}^{(n)})\tilde p(X_{i,t}^{(n)}|X_{<t}^{(n)})} \Big] \frac{p(Y_{i,1}|X_{i,1}^{(n)})p(X_{i,1}^{(n)}|\theta^{(n)})}{\tilde p(Y_{i,1}|X_{i,1}^{(n)}) \tilde p(X_{i,1}^{(n)}|\theta^{(n)})} \Big \} \\
    &= \Big \{\prod_{i=1}^m \frac{p(Y_{i,T}|X_{i,T}^{(n)})p(X_{i,T}^{(n)}|X_{<T}^{(n)})}{\tilde p(Y_{i,T}|X_{i,T}^{(n)})\tilde p(X_{i,T}^{(n)}|X_{<T}^{(n)})}\Big \} \\
    &\quad\quad\quad \times \Big \{ \prod_{i=1}^m \Big [\prod_{t=2}^{T-1} \frac{p(Y_{i,t}|X_{i,t}^{(n)})p(X_{i,t}^{(n)}|X_{<t}^{(n)})}{\tilde p(Y_{i,t}|X_{i,t}^{(n)})\tilde p(X_{i,t}^{(n)}|X_{<t}^{(n)})} \Big] \frac{p(Y_{i,1}|X_{i,1}^{(n)})p(X_{i,1}^{(n)}|\theta^{(n)})}{\tilde p(Y_{i,1}|X_{i,1}^{(n)}) \tilde p(X_{i,1}^{(n)}|\theta^{(n)})} \Big \} \\
    &= \Big \{\prod_{i=1}^m \frac{p(Y_{i,T}|X_{i,T}^{(n)})p(X_{i,T}^{(n)}|X_{<T}^{(n)})}{\tilde p(Y_{i,T}|X_{i,T}^{(n)})\tilde p(X_{i,T}^{(n)}|X_{<T}^{(n)})}\Big \} w_{T-1}^{(n)},
\end{align*}
with $w_1^{(n)} = \prod_{i=1}^m \frac{p(Y_{i,1}|X_{i,1}^{(n)})p(X_{i,1}^{(n)}|\theta^{(n)})}{\tilde p(Y_{i,1}|X_{i,1}^{(n)}) \tilde p(X_{i,1}^{(n)}|\theta^{(n)})}$. Then, a resampling procedure is applied, where at each time step, should the effective sample size, $N_{\text{eff}} = \big[\sum_{n=1}^N (\tilde w_t^{(n)})^2\big]^{-1}$, fall below some threshold, $N_{\text{thresh}}$, we sample with replacement from the sample trajectories with probabilities determined by the normalized weights at the current time step \citep{doucet2000}.

\subsection{Weight Formulation}

\begin{figure}
    \centering
    \includegraphics[width=\linewidth]{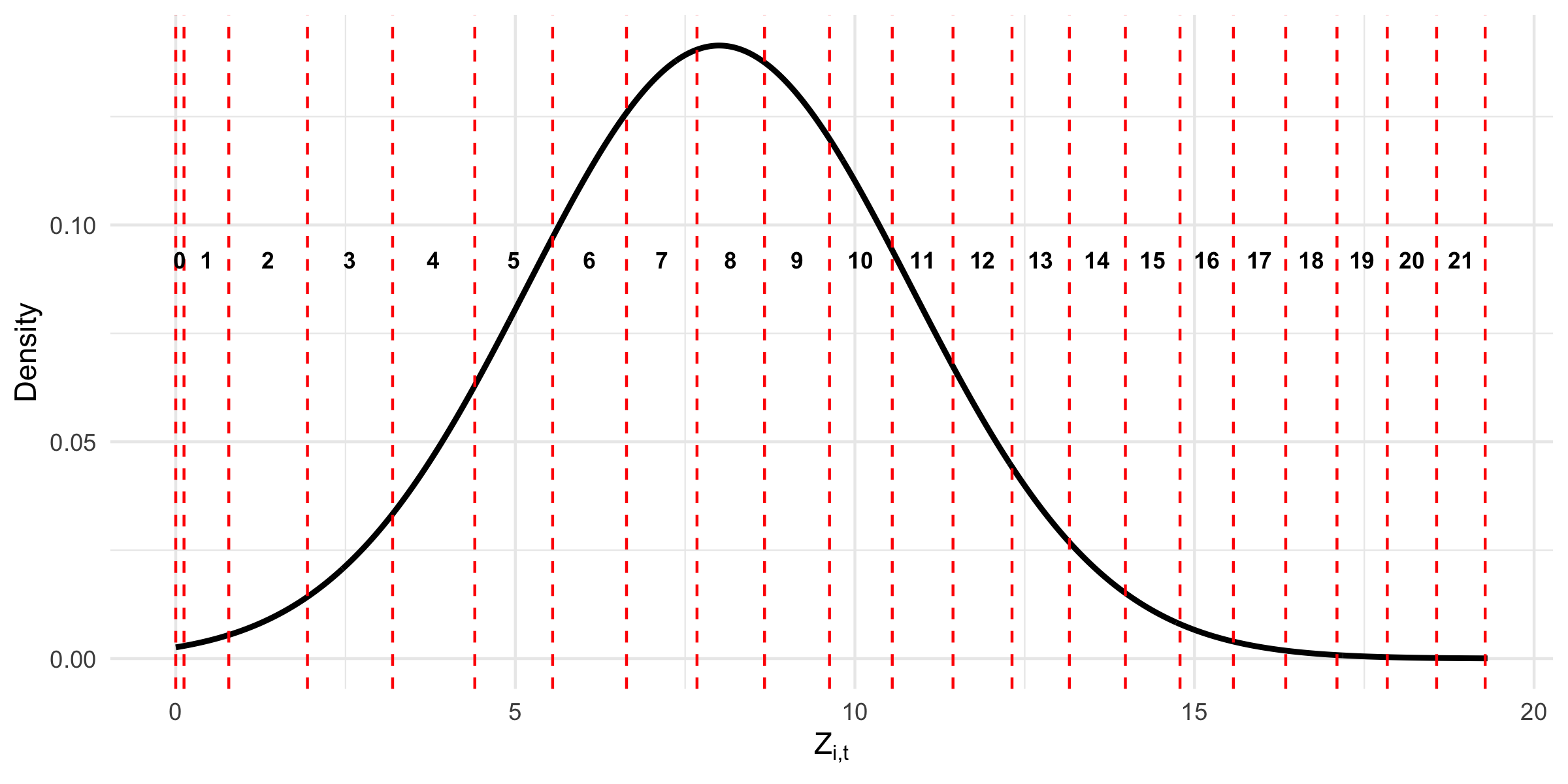}
    \caption{Density of $Z_{i,t}$ ($\text{N}_{(0,\infty)}(8,8)$) with superimposed lines corresponding to the intervals, $A_{0}$, $A_1$, ..., $A_{21}$, that map to counts $X_{i,t} = 0$, $X_{i,t} = 1$, ..., $X_{i,t} = 21$ with probabilities defined by Poisson($\lambda = 8$).}
    \label{LGMAP_fig}
\end{figure}

Notice that the denominator of the importance weights requires the discrete random variables $\{X_{i,t}\}$ when the proposal posterior in \eqref{LGPNAR} is defined for the Gaussian random variables $\{Z_{i,t}\}$. However, as previously detailed, the mapping \eqref{LGMap} recovers random variables with the desired count distribution. This is done by partitioning the domain of $Z_{i,t}$ into intervals, $A_{X_{i,t}}$, that are the preimage of each integer $X_{i,t}$ under the mapping, visualized in \autoref{LGMAP_fig}, and defined as
\begin{equation}\label{AXT}
    \begin{aligned}
        A_{X_{i,t}} &:= \Big (\tilde\lambda_{i,t} + \sqrt{\tilde\lambda_{i,t}} \Phi_{(-\sqrt{\tilde\lambda_{i,t}},\infty)}^{-1}[F_{\tilde\lambda_{i,t}}(X_{i,t} - 1)], \tilde\lambda_{i,t} + \sqrt{\tilde\lambda_{i,t}} \Phi_{(-\sqrt{\tilde\lambda_{i,t}},\infty)}^{-1}[F_{\tilde\lambda_t}(X_{i,t})]\Big ] \\
        &= (a_{X_{i,t}},b_{X_{i,t}}].
    \end{aligned}
\end{equation}
Thus, for the mapping to have recovered a sample $X_{i,t}$, any of the $Z_{i,t} \in A_{X_{i,t}}$ may have been proposed by the surrogate model. The equivalent expressions to the terms $\tilde p(Y_{i,t}|X_{i,t})$ and $\tilde p(X_{i,t}|X_{<t})$ in the denominator of the importance weights must then be $\tilde p(Y_{i,t}|Z_{i,t} \in A_{X_{i,t}})$ and $\tilde p(Z_{i,t} \in A_{X_{i,t}}|Z_{<t})$, which are the probability of observing a count of $Y_{i,t}$ given the true count is $X_{i,t}$, and the probability of proposing a $Z_{i,t}$ that maps to $X_{i,t}$ under the surrogate model. This second term, $\tilde p(Z_{i,t} \in A_{X_{i,t}} | Z_{<t})$ has a closed form solution as follows:
    \begin{align}\label{closedform}
    \tilde p(Z_{i,t} \in A_{X_{i,t}} | Z_{<t}) &= \int_{A_{X_{i,t}}} \tilde p(Z_{i,t} | Z_{<t}) dZ_{i,t} \notag \\
    &= \int_{a_{X_{i,t}}}^{b_{X_{i,t}}} \text{N}_{(0,\infty)}(\tilde\lambda_{i,t},\tilde\lambda_{i,t}) dZ_{i,t} \notag \\
    &= \int_{\frac{a_{X_{i,t}}-\tilde\lambda_{i,t}}{\sqrt{\tilde\lambda_{i,t}}}}^{\frac{b_{X_{i,t}}-\tilde\lambda_{i,t}}{\sqrt{\tilde\lambda_{i,t}}}} \text{N}_{(-\sqrt{\tilde\lambda_{i,t}},\infty)}(0,1) dZ_{i,t} \notag \\
    &= \Phi_{(-\sqrt{\tilde\lambda_{i,t}},\infty)}\{\Phi_{(-\sqrt{\tilde\lambda_{i,t}},\infty)}^{-1}[F_{\tilde\lambda_t}(X_{i,t})]\} \notag \\
    &\quad\quad- \Phi_{(-\sqrt{\tilde\lambda_{i,t}},\infty)}\{\Phi_{(-\sqrt{\tilde\lambda_{i,t}},\infty)}^{-1}[F_{\tilde\lambda_t}(X_{i,t} -1)]\} \notag \\
    &= F_{\tilde\lambda_t}(X_{i,t}) - F_{\tilde\lambda_t}(X_{i,t} -1) \notag \\
    &= f_{\tilde\lambda_t}(X_{i,t}),
    \end{align}
where if $F(\cdot)$ corresponds to a Poisson cumulative distribution function, then $\tilde p(Z_{i,t} \in A_{X_{i,t}} | Z_{<t}) = \text{Pois}(\tilde\lambda_{i,t})$. This identifies that the probability of getting any $Z_{i,t}$ from a truncated normal density with mean and variance $\tilde\lambda_{i,t}$ that maps to a chosen $X_{i,t}$ with a Poisson marginal distribution is exactly the probability of getting $X_{i,t}$ from a Poisson with rate $\tilde\lambda_{i,t}$. A near-identical argument can be made in the absence of truncation or under a different count distribution, provided an appropriate normal approximation to that count distribution is chosen. 

However, the term $\tilde p(Y_{i,t}|Z_{i,t} \in A_{X_{i,t}})$ has no closed-form solution, so we consider two different approximations for the denominator of $w_t$, providing two different weight schemes. The first uses \eqref{closedform} and a simple approximation of the form
\begin{equation}\label{ytgivenxt}
    \begin{aligned}
        \tilde p(Y_{i,t}|Z_{i,t} \in A_{X_{i,t}}) &\approx \sum_{g=1}^G \tilde p(Y_{i,t}|Z_{i,t} = z_g) \tilde p(Z_{i,t} = z_g | Z_{i,t} \in A_{X_{i,t}}, Z_{<t}) \Delta z_g \\
        &\approx \frac{1}{G}\sum_{g=1}^G \tilde p(Y_{i,t}|Z_{i,t} = z_g),
    \end{aligned} 
\end{equation}
where $z_1,z_2,...z_G$ are a grid of $G$ equally spaced values of $Z_{i,t}$ across the interval $A_{X_{i,t}}$, so $\Delta z = \frac{b_{X_{i,t}}-a_{X_{i,t}}}{G}$ for each $z_g$. This approach averages $\tilde p(Y_{i,t}|Z_{i,t} = z_g)$ for each $z_g$ across the interval, and thus, assumes the uniformity of $\tilde p(Z_{i,t}|Z_{<t})$ across the set $A_{X_{i,t}}$. Accordingly, the weight scheme using this approximation will henceforth be referred to as ``grid-based". 

The other weight scheme provides an approximation for the product of the two terms by jointly integrating with respect to $Z_{i,t}$ over the set $A_{X_{i,t}}$
\begin{align}\label{intdenom}
    \tilde p(Y_{i,t}|Z_{i,t} \in A_{X_{i,t}}) \tilde p(Z_{i,t} \in A_{X_{i,t}}|Z_{<t}) &= \tilde p(Y_{i,t},Z_{i,t}\in A_{X_{i,t}}|Z_{<t}) \notag \\
    &= \int_{A_{X_{i,t}}} \tilde p(Y_{i,t}|Z_{i,t})\tilde p(Z_{i,t}|Z_{<t})dZ_{i,t} \notag \\
    &= \int_{A_{X_{i,t}}} f(Z_{i,t})dZ_{i,t},
\end{align}
where the $f(Z_{i,t})$ indicates that the integrand is merely a function of $Z_{i,t}$. This integral also has no closed-form solution, so we suggest using Gauss-Legendre quadrature, where instead of choosing uniformly spaced points as in \eqref{ytgivenxt}, we consider choosing points defined by the roots of Legendre polynomials. Such a procedure requires the bounds of integration to be $(-1,1)$, and so, to map from $(a_{X_{i,t}},b_{X_{i,t}}]$ to $(-1,1]$, apply the transformation $q(Z_{i,t}) = \frac{2}{b_{X_{i,t}}-a_{X_{i,t}}}Z_{i,t} - \frac{a_{X_{i,t}}+b_{X_{i,t}}}{b_{X_{i,t}}-a_{X_{i,t}}}$. Then by change of variables and Gauss-Legendre quadrature,
\begin{align}\label{quadrature}
    p(Y_{i,t}|Z_{i,t} \in A_{X_{i,t}})p(Z_{i,t} \in A_{X_{i,t}}|Z_{<t}) &= \int_{-1}^1 f\Big(\frac{b_{X_{i,t}}-a_{X_{i,t}}}{2} q + \frac{a_{X_{i,t}}+b_{X_{i,t}}}{2}\Big)\frac{b_{X_{i,t}}-a_{X_{i,t}}}{2}dq \notag \\ 
            &\approx \frac{b_{X_{i,t}}-a_{X_{i,t}}}{2} \sum_{j=1}^Q w^q_j f\Big(\frac{b_{X_{i,t}}-a_{X_{i,t}}}{2}q_j + \frac{a_{X_{i,t}}+b_{X_{i,t}}}{2}\Big),
\end{align}
where $q_j$ is the $j^{th}$ root of the $Q^{th}$ Legendre polynomial, $w^q_j = \frac{2}{(1-q_j^2)[P_Q'(q_j)]^2}$ is the quadrature weight, and $P_Q'$ is the first derivative of the $Q^{th}$ Legendre polynomial. Using such an approach will henceforth be referred to as ``quadrature-based".

Our proposed modelling framework for conducting inference under a desired thinned count autoregression, using the surrogate model as a proposal posterior and either of the importance weight schemes, is summarized in \autoref{algo}.

\begin{algorithm}[]\caption{Inference for thinned count autoregressions}\label{algo}
    Fit surrogate model \eqref{LGPNAR} using HMC, obtaining samples $Z_{i,t}^{(n)}$ and $\theta^{(n)}$ for all $i=1$ \KwTo $m$, $t=1$ \KwTo $T$, $n=1$ \KwTo $N$;

    Map from $\displaystyle Z_{i,t}^{(n)}$ to $X_{i,t}^{(n)} = F_{\tilde\lambda_{i,t}^{(n)}}^{-1}\Big [\Phi_{(-\sqrt{\tilde\lambda_{i,t}^{(n)}},\infty)}\Big(\frac{Z_{i,t}^{(n)} - \tilde\lambda_{i,t}^{(n)}}{\sqrt{\tilde\lambda_{i,t}^{(n)}}}\Big) \Big]$ for all $i=1$ \KwTo $m$, $t=1$ \KwTo $T$, $n=1$ \KwTo $N$;

    Calculate initial weights $w^{(n)}_1 = \prod_{i=1}^m \frac{p\big(Y_{i,1}|X_{i,1}^{(n)},\theta^{(n)}\big)p\big(X_{i,1}^{(n)}|\theta^{(n)}\big)}{\tilde p\big(Y_{i,1}|Z_{i,1}^{(n)} \in A_{X_{i,1}^{(n)}},\theta^{(n)}\big)\tilde p\big(Z_{i,1}^{(n)} \in A_{X_{i,1}^{(n)}} |\theta^{(n)}\big)}$ using selected weight scheme for all $n=1$ \KwTo $N$;

    \For{$t = 2$ \KwTo $T$}{
        Calculate weights $w^{(n)}_t = w^{(n)}_{t-1} \prod_{i=1}^m\frac{p\big(Y_{i,t}|X_{i,t}^{(n)},\theta^{(n)}\big)p\big(X_{i,t}^{(n)}|X_{<t}^{(n)},\theta^{(n)}\big)}{\tilde p\big(Y_{i,t}|Z_{i,t}^{(n)} \in A_{X_{i,t}^{(n)}},\theta^{(n)}\big)\tilde p\big(Z_{i,t}^{(n)} \in A_{X_{i,t}^{(n)}}|Z_{<t}^{(n)},\theta^{(n)}\big)}$ using selected weight scheme for all $n=1$ \KwTo $N$;

        Normalize the weights $\tilde w^{(n)}_t = \frac{w_t^{(n)}}{\sum_{k=1}^N w_t^{(k)}}$ for all $n=1$ \KwTo $N$;

        Calculate effective sample size $N_{\text{eff}} = \big[\sum_{n=1}^N (\tilde w_t^{(n)})^2\big]^{-1}$;

        \If{$N_{\text{eff}} < N_{\text{thresh}}$}{
            Sample indices $j^{(n)}$ from the probability distribution $\mathbbm{P}(j=n) = \tilde w_t^{(n)}$ for all $n=1$ \KwTo $N$;

            Replace $Z_{\cdot,1:T}^{(n)}$, $X_{\cdot,1:T}^{(n)}$, and $\theta^{(n)}$ with $Z_{\cdot,1:T}^{j^{(n)}}$, $X_{\cdot,1:T}^{j^{(n)}}$, and $\theta^{j^{(n)}}$, set $w_t^{(n)} = 1/N$ for all $n=1$ \KwTo $N$;
        }
    }

    Normalize the weights $\tilde w^{(n)}_T = \frac{w_T^{(n)}}{\sum_{k=1}^N w_T^{(k)}}$ for all $n=1$ \KwTo $N$;

    Compute desired expectation $\mathbbm{E}_{p}[h(X,\theta)] \approx \sum_{n=1}^N h(X^{(n)},\theta^{(n)})\tilde w_T^{(n)}$;
\end{algorithm}

\section{Simulation study}\label{sec:simstudy}
\subsection{Simulation setup}
To demonstrate the feasibility, accuracy, and validity of the importance sampling smoother, we perform a simulation study comparing estimates under the surrogate posterior and under both importance weight schemes to estimates under the ``true" posterior for a simple thinned count autoregression.

The simplest example of a BTPNAR with autoregression of order 1 can be fit using the slice sampler \citep{neal2003} for $\{X_t\}$ and a random-walk sampler for the remaining parameters, implemented in the R software package Nimble \citep{nimble}, provided the time series is not too long. In this study, for $T=50$, the maximum observed $\hat R$ value is approximately $1.015$. When averaged across all replications and parameter combinations, the maximum $\hat R$ value decreases to approximately $1.002$, indicating strong posterior convergence. The procedure that guarantees this convergence is slow, as Markov chains must be run for a long time to obtain a sufficient effective sample size since the high-dimensional, autocorrelated structure of $\{X_t\}$ causes poor mixing. It is challenging to model longer time series, with or without additional model complexity, as obtaining posterior convergence is inconsistent \citep{bracher2021, slatermobility}. Such a model that may be fit directly is defined as the left-hand side below:
\begin{align}
        \begin{aligned}
            Y_t | X_t &\sim \text{Bin}(X_t,\pi) \\
            X_t | X_{t-1} &\sim \text{Pois}(\lambda_t) \quad \text{for }t\geq 2 \\
            \lambda_t &= \nu + \phi X_{t-1} \\
            \nu &\sim \text{N}_{(0,\infty)}(\nu_0,5^2) \\
            \phi &\sim \text{Unif}(0,1) \\
            \pi &\sim \text{N}_{(0,1)}(0.6,0.3^2) \\
            X_1 &= x_1, \\
        \end{aligned}
\quad
&
\quad
        \begin{aligned}
            Y_t | Z_t &\sim \text{N}(\pi Z_t, \pi(1-\pi)Z_t) \\
            Z_t | Z_{t-1} &\sim \text{N}_{(0.1,\infty)}(\tilde\lambda_t,\tilde\lambda_t) \quad \text{for }t\geq 2 \\
            \tilde\lambda_t &= \nu + \phi Z_{t-1} \\
            \nu &\sim \text{N}_{(0,\infty)}(\nu_0,5^2) \\
            \phi &\sim \text{Unif}(0,1) \\
            \pi &\sim \text{N}_{(0,1)}(0.6,0.3^2) \\
            Z_1 &= \text{midpoint}(A_{{x_1}}), \\
        \end{aligned}\label{simModels}
\end{align}
with $X_t = F_{\tilde\lambda_{t}}^{-1}\Big [\Phi_{\Big(\frac{0.1 - \tilde\lambda_t}{\sqrt{\tilde\lambda_t}},\infty\Big)}\Big(\frac{Z_{t} - \tilde\lambda_{t}}{\sqrt{\tilde\lambda_{t}}}\Big) \Big]$. The surrogate model to this posterior, defined by the right-hand side model of \eqref{simModels}, can be fit with more ease using the No-U-Turn variant of HMC implemented in Rstan \citep{stan}. This surrogate, ignoring the truncation on $\tilde p(Z_t|Z_{t-1})$, is defined so that it has the same conditional moments as the true posterior and the same or analogous priors, under the guidance of \citet{slater2025}. An example of an analogous prior notes that, where in the true posterior, we fix $X_1 = x_1$ as a known quantity to improve the stability of our estimates, the analogous paradigm for the surrogate model fixes a single value of $Z_1 \in A_{x_1}$, as any value in that set maps to $x_1$. Accordingly, we chose the midpoint as the single representative of this interval. We note that the lower bound of $\tilde p(Z_t|Z_{t-1})$ is now $0.1$ as smaller values cause the density $\tilde p(Y_t|Z_t)$ to approach a point mass around its mean, such that only one value of $Y_t$ will have a non-zero density, thereby making our importance weights ill-defined. 

50 replications of time series of length $T=50$ are simulated using the formulation described by \eqref{PAR} and \eqref{BT} for each of $24$ different combinations of the parameters $(\nu,\phi,\pi)$ chosen to induce $\{Y_t\}$ to have means varying from $3.33$ to $41.67$. Specifically, either $\nu = 5,10$ and $\phi$ was chosen to allow for the calculated value of $\pi = \mathbbm{E}[Y_t](1-\phi)/\nu$ to either be ``small" or ``large" alongside or opposite $\phi$. This form for $\pi$ comes solving $\mathbbm{E}[X_t] = \frac{\nu}{1-\phi}$ and $\mathbbm{E}[Y_t] = \pi \mathbbm{E}[X_t]$. Accordingly, we can see performance in instances with lower means, resulting in the surrogate model performing poorly, and with varying levels of underreporting. For each of these chosen $\nu$, we set $\nu_0 = \nu$ in the prior. 

The true posterior, induced by the left-hand side of \eqref{simModels}, is sampled from using Nimble with four chains in parallel for 260,000 iterations, discarding the first 10,000 as warmup, and keeping only every tenth sample to guarantee a large effective sample size, necessary to function as the ``truth" in our comparison. We follow the first two steps of \autoref{algo} in fitting the surrogate posterior, induced by the right-hand side model in \eqref{simModels}, using Rstan with four chains in parallel for 7000 iterations, discarding the first 3000 as warmup, and mapping samples of $\{Z_{t}\}$ to $\{X_t\}$ using \eqref{LGMap}. For both importance weight schemes, the surrogate model is treated as the proposal posterior and $\tilde w_T^{(n)}$ is calculated using all but the final step of \autoref{algo}.

We must now consider how to compare inference conducted using the posterior samples from the true model, the unadjusted surrogate model, and from the surrogate when adjusted with either of the importance weight schemes. For this, we propose comparing the correspondence in estimated medians and 95\% credible intervals, using two metrics: the ``perfect match rate" and the ``distance to perfect match". The former is the proportion of lower, median, and upper quantiles calculated under the surrogate or adjusted models that exactly match those of the true model, with a perfect match rate of $100\%$ indicating perfect correspondence. As an example, for a time series with $T=2$, should the true estimated quantiles of $X_1$ and $X_2$ be $(1,3,7)$ and $(2,4,7)$, and should the surrogate model estimate the quantiles as $(1,3,5)$ and $(2,6,7)$, we would compute a 67\% perfect match rate. The latter metric notes that there will be instances in which the importance sampling adjustments may improve estimates without providing a perfect match. Thus, we also consider the absolute distance between the quantiles computed under the true model and both the surrogate and adjusted models. A distance to perfect match of $0$ indicates the estimated values are equivalent, and therefore also coincides with a perfect match rate of $100\%$. Continuing with the same example, should one of the weight schemes provide the estimates $(1,3,6)$ and $(2,5,7)$, there would still be a 67\% perfect match rate, but the distance to perfect match would reduce from 4 to 2. We accordingly calculate estimates for medians and $95\%$ credible intervals for $\{X_t\}$ under both models in \eqref{simModels} and under the adjustments made by both importance sampling weight schemes, using $G=Q=10$ in \eqref{ytgivenxt} and \eqref{quadrature}. The estimate of the lower 2.5\% quantile, for example, may be computed as $\tilde X_{t_{0.025}} \approx \min\{x \in X_{t_{(1)}},...,X_{t_{(N)}}: \sum_{n=1}^N \tilde w^{(n)}_T \mathbbm{1}\{X^{(n)}_t \leq x\} \geq 0.025 \}$, where $X_{t_{(k)}}$ refers to the $k^{th}$ smallest sample of $X_t$.

\subsection{Simulation results}

\begin{figure}[]
    \centering
    \begin{subfigure}{\textwidth}
        \centering
        \includegraphics[width=\textwidth]{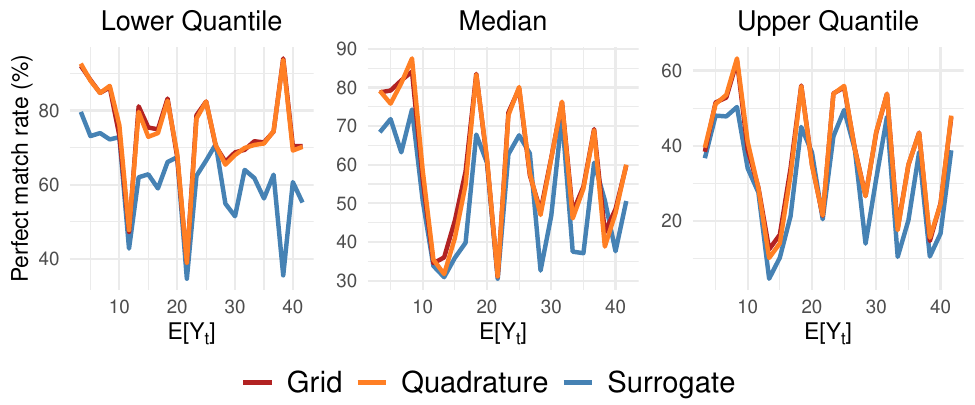}
        \caption{}
    \end{subfigure}
    \begin{subfigure}{\textwidth}
        \centering
        \includegraphics[width=\textwidth]{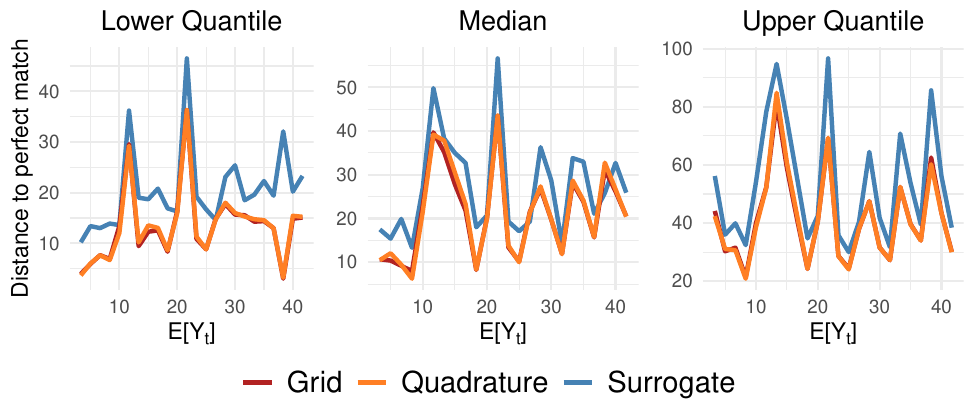}
        \caption{}
    \end{subfigure}
    \caption{(a) Perfect match rates and (b) distance to perfect matches for estimated medians and $95\%$ credible intervals for the approximate model, and under both grid-based and quadrature-based weight schemes.}
    \label{matchrates_fig}
\end{figure}

The calculated metrics for the lower, median, and upper quantiles for each of the 24 parameter combinations are summarized in \autoref{matchrates_fig}, \autoref{perfect_match_rates}, and \autoref{dist_match_rates}, where each entry is found by averaging the results over the 50 corresponding replications. Both of the importance weight schemes are effective on average, as we observe an increase in the perfect match rate and a decrease in the distance to perfect matches. Perfect match rates for the lower quantile increases from $61.2\%$ to $74.2\%$ under the grid-based scheme and to $74.0\%$ under the quadrature-based scheme. On the median, we observe a $8.4\%$ and $7.6\%$ increase for either scheme, and increases of $6.7\%$ and $6.6\%$ for the upper quantile. On average across all simulated time series, the distances to perfect match for the surrogate model are $20.5$, $27.1$, and $53.6$. Our grid-based adjustments removed $34.9\%$, $23.5\%$, and $21.8\%$ of this error, and similarly, $34.7\%$, $22.6\%$, and $22.0\%$ of the total error was removed with quadrature-based adjustments. The grid-based scheme thus slightly outperformed the quadrature-based scheme for both metrics.
Notably, in some instances, the importance sampling adjustments decrease accuracy of the estimated quantiles (see rows corresponding to $\mathbbm{E}[Y_t] = 20,26.67,38.33$ in \autoref{perfect_match_rates}). \autoref{dist_match_rates} indicates that the severity of this ``lack" of perfect matches is small, as in the worst case, on the median of $\mathbbm{E}[Y_t] = 38.33$, we add an average distance of $6.966$. In other words, in the worst-case scenario, the importance weights moved each median estimate of $X_t$ away from the truth by approximately 0.139. This is likely due to simulation noise, as there is numerical error when fitting both posteriors, computing the importance weights, and computing the estimated expectations. In \autoref{matchrates_fig}, the perfect match rates appear to decrease as $\mathbbm{E}[Y_t]$ increases, likely because as the mean increases, more candidate values of $X_t$ are similarly probable. However, the rate of this trend is slower than that of $\mathbbm{E}[Y_t]$ itself. A similar phenomenon is visible in the distance to perfect matches, as they increase with $\mathbbm{E}[Y_t]$, but at a slower rate than $\mathbbm{E}[Y_t]$. In both instances, this indicates that the surrogate framework improves as the means of the time series grow, agreeing with our prior assumptions.

These results indicate that the importance weights are effective in improving our estimates of medians and 95\% credible intervals. As the grid-based scheme performs best, it will be used for the case studies in \autoref{sec:casestudies}.

\section{Case studies}\label{sec:casestudies}

\subsection{Univariate: rotavirus in Germany}

\begin{figure}[]
    \centering
    \begin{subfigure}{\textwidth}
        \centering
        \includegraphics[width=\textwidth]{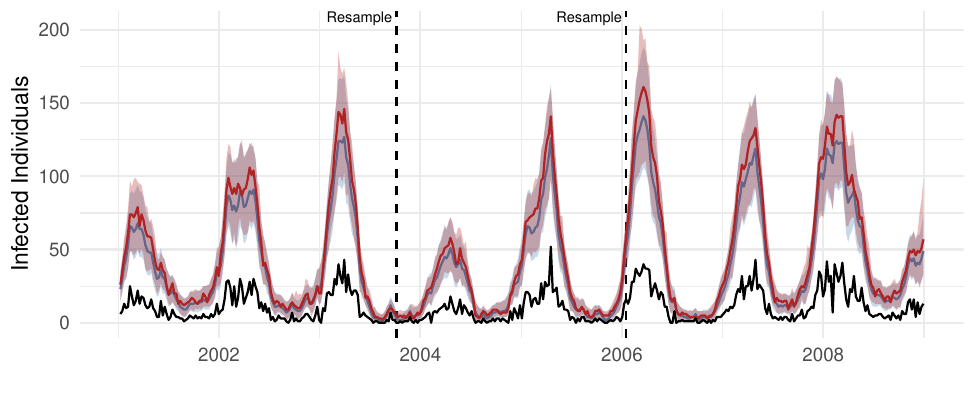}
        \caption{Reconstructed and reweighted incidence}
    \end{subfigure}
    \begin{subfigure}{0.32\textwidth}
        \centering
        \includegraphics[width=\textwidth]{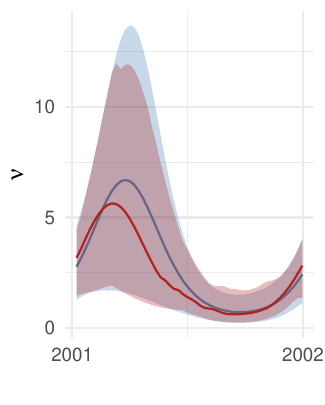}
        \caption{Endemic component}
    \end{subfigure}
    \hfill
    \begin{subfigure}{0.32\textwidth}
        \centering
        \includegraphics[width=\textwidth]{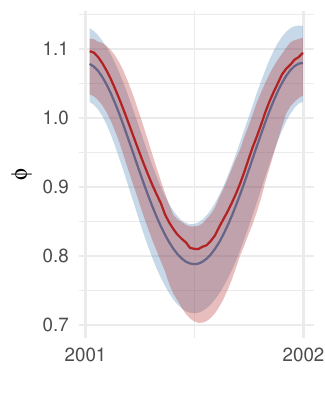}
        \caption{Reproductive rate}
    \end{subfigure}
    \hfill
    \begin{subfigure}{0.32\textwidth}
        \centering
        \includegraphics[width=\textwidth]{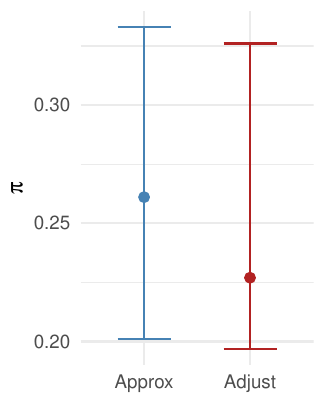}
        \caption{Reporting probability}
    \end{subfigure}
    \begin{subfigure}{\textwidth}
        \centering
        \includegraphics[width=\textwidth]{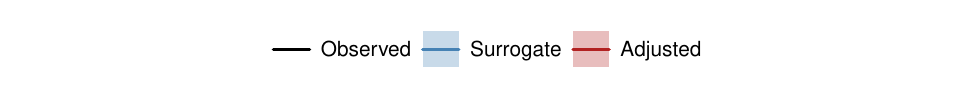}
    \end{subfigure}

    \caption{a) Observed, reconstructed, and reweighted rotavirus counts in 2001-2008 Saarland, Germany. Estimated and reweighted b) cases not attributable to past cases, c) reproductive rate, d) reporting probability. Since both $\nu$ and $\phi$ are only functions of $t$ and constants, only the first period (52 weeks) is shown. All with posterior and reweighted medians and $95\%$ credible intervals.}
    \label{saarland_fig}
\end{figure}

With approximately 95\% of children worldwide infected by age 5, rotavirus is the prevailing cause of severe diarrhea in infants. Other symptoms of rotavirus include vomiting and fever, and the disease is typically transmitted through fecal-to-oral spread \citep{bernstein2009}. This case study analyzes weekly rotavirus data from Germany during the time period 2001-2008 that was previously studied in \citet{weidemann}, \citet{bracher2021}, and \citet{slater2025}. For our surrogate model, we take inspiration from \citet{slater2025} to illustrate the applicability of our importance sampling methodology to models more complicated than those used in \autoref{sec:simstudy}:
\begin{equation}\label{saarland}
    \begin{aligned}
        Y_t|Z_t &\sim \text{N}(\pi Z_t, \pi(1-\pi)Z_t) \\
        Z_t|Z_{t-1} &\sim \text{N}_{(0.1,\infty)}(\tilde\lambda_t,\tilde\lambda_t) \\
        \tilde\lambda_t &= \nu_t + \phi_t Z_{t-1} \quad \text{for }t\geq 2 \\
        \log(\nu_t) &= \alpha^{(\nu)} + \gamma_1^{(\nu)}\sin\Big(\frac{2\pi t}{52}\Big) + \gamma_2^{(\nu)}\cos\Big(\frac{2\pi t}{52}\Big) \\
        \log(\phi_t) &= \alpha^{(\phi)} + \gamma_1^{(\phi)}\sin\Big(\frac{2\pi t}{52}\Big) + \gamma_2^{(\phi)}\cos\Big(\frac{2\pi t}{52}\Big) \\
        \tilde\lambda_1 &\sim \text{N}_{(0,\infty)}(10,10^2) \\
        \alpha^{(\nu)},\alpha^{(\phi)},\gamma_1^{(\nu)},\gamma_1^{(\phi)},\gamma_2^{(\nu)},\gamma_2^{(\phi)} &\sim \text{N}(0,1) \\
        \text{logit}(\pi) &\sim \text{N}(0,2^2),
    \end{aligned}
\end{equation}
then apply the rest of \autoref{algo} to conduct inference. The top few lines of this model are the same as the surrogate model in \ref{simModels}, but the key complications are the seasonal expansions of $\nu_t$ and $\phi_t$. The reconstructed incidence is visualized in \autoref{saarland_fig}, where the black line corresponds to the observed case counts $\{Y_t\}$, the blue ribbons and line correspond to 95\% credible intervals and medians for the unknown true case counts $\{X_t\}$ recovered from mapping $\{Z_t\}$, and the red ribbons and line correspond to 95\% credible intervals and medians for the estimates found from importance sampling. We also show medians and 95\% credible intervals estimates for one year for both $\nu$ and $\phi$ using the same colouring. Resampling occurs only twice, at $t=144,263$, indicating that the surrogate model is a well-suited proposal, as the effective sample sizes decline slowly (see \autoref{saarlandESS_fig}). The importance sampling weights pull the estimated quantiles of $\{X_t\}$ up, lower the magnitude of each spike in the endemic component, slightly increase estimates of $\phi$, and decrease the median of the reporting probability from $26.10\%$ to $22.69$\%. On the scale of the observed cases, these adjustments to the true case counts are substantial. \autoref{saarlandCHANGE_fig} shows that the magnitude of adjustments to the median is 56.5\% of the magnitude of the observed number of cases. This means, for the observed time series $\{Y_t\}$ with $\mathbbm{E}[Y_t] = 10.46$, the median estimates change by $5.91$, on average. These changes appear to fluctuate alongside fluctuations in $\{Y_t\}$, where large spikes in $\{Y_t\}$ coincide with large spikes in the adjustments to $\{X_t\}$. The average adjustments are all positive, indicating that estimates for all the lower, median, and upper quantiles are raised, but the upper quantile estimates are regularly lowered. These results indicate that fitting the unadjusted surrogate model leads to underestimates of incidence and prevalence.

\subsection{Multivariate: meningococcal infections in France}

\begin{figure}[]
    \centering

    \begin{subfigure}{\textwidth}
        \centering
        \includegraphics[width=\textwidth]{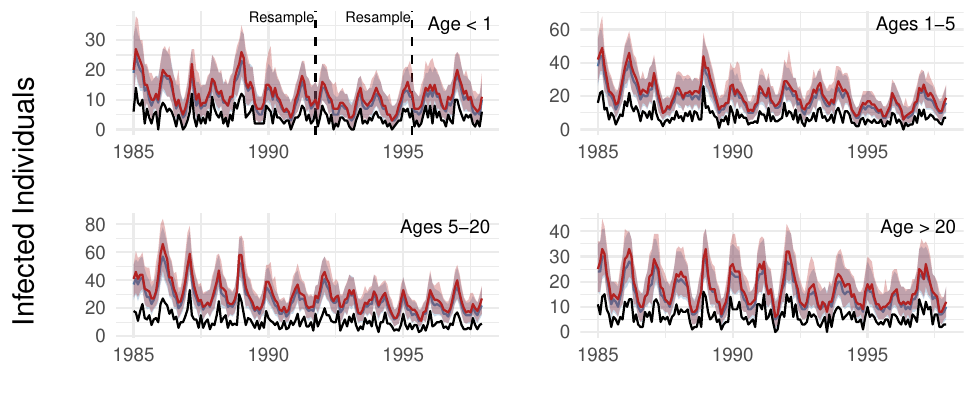}
        \caption{Reconstructed and reweighted incidence}
    \end{subfigure}

    \begin{subfigure}{\textwidth}
        \centering
        \includegraphics[width=\textwidth]{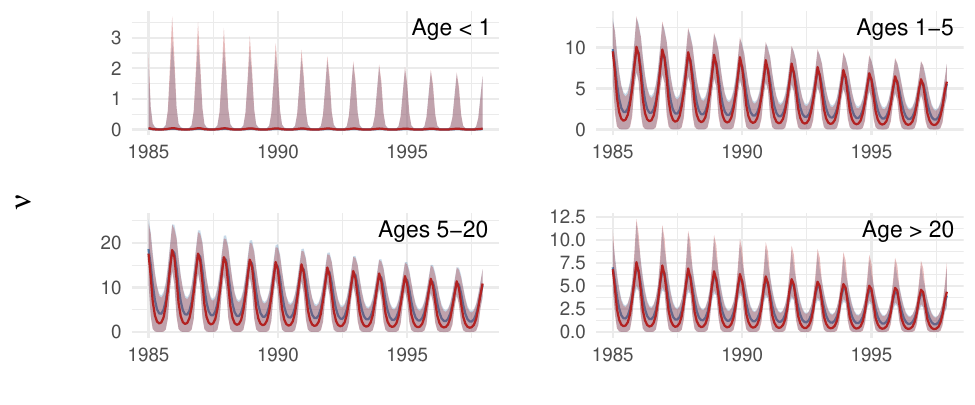}
        \caption{Cases not attributable to past cases or other age groups}
    \end{subfigure}

    \begin{subfigure}{0.32\textwidth}
        \centering
        \includegraphics[width=\textwidth]{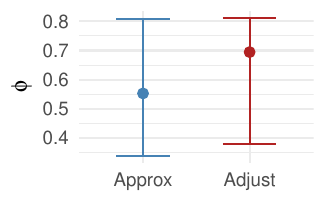}
        \caption{Within-group reproduction}
    \end{subfigure}
    \hfill
    \begin{subfigure}{0.32\textwidth}
        \centering
        \includegraphics[width=\textwidth]{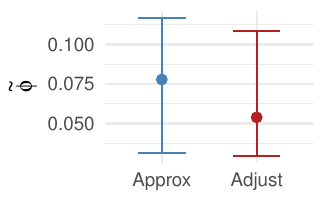}
        \caption{Out-of-group reproduction}
    \end{subfigure}
    \hfill
    \begin{subfigure}{0.32\textwidth}
        \centering
        \includegraphics[width=\textwidth]{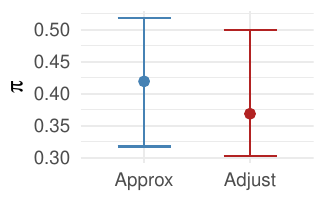}
        \caption{Reporting probability}
    \end{subfigure}

    \begin{subfigure}{\textwidth}
        \centering
        \includegraphics[width=\textwidth]{LEGEND.pdf}
    \end{subfigure}

    \caption{a) Observed, reconstructed, and reweighted meningococcal counts in 1985-1997 France for each age group. Estimated and reweighted b) endemic component for each age group, c) within-group reproductive rate, d) out-of-group reproductive rate, and e) reporting probability. All with posterior and reweighted medians and $95\%$ credible intervals.}
    \label{france_fig}
\end{figure}

Meningococcus infects between 500,000 and 1.2 million people yearly, with roughly a 10\% mortality rate. These infections may present as meningitis with septicemia, pneumonia, pericarditis, septic arthritis, conjunctivitis, and urethritis. Transmission is solely between humans, usually occurring through respiratory droplets, with children being the most susceptible prior to developing antibodies \citep{stephens2012}. This case study analyzes monthly meningococcal incidence in France across $m=4$ age groups during the period of 1985-1997 as studied in \citet{held2005}. We use the below normal-normal approximation with $\tilde\lambda_{i,t}$ and $\log(\nu_{i,t})$ inspired by their model formulation as our surrogate model:
\begin{equation}\label{france}
    \begin{aligned}
            Y_{i,t}|Z_{i,t} &\sim \text{N}(\pi Z_{i,t}, \pi(1-\pi)Z_{i,t}) \\
            Z_{i,t}|Z_{<t} &\sim \text{N}_{(0.1,\infty)}(\tilde\lambda_{i,t},\tilde\lambda_{i,t}) \\
            \tilde\lambda_{i,t} &= n_{i,t}\nu_{i,t} + \phi Z_{i,t-1} + \tilde\phi \sum_{j\neq i} Z_{j,t-1} \quad \text{for }t\geq 2\\
            \log(\nu_{i,t}) &= \alpha_i + \beta t + \gamma_1 \sin \Big(\frac{2\pi t}{12}\Big) + \gamma_2\cos\Big(\frac{2\pi t}{12}\Big) \\
            \tilde\lambda_{i,1} &\sim \text{N}_{(0,\infty)}(10,10^2) \quad \text{for } i = 1,...,m \\
            \tilde\phi &\sim \text{Unif}(0,1/(m-1)) \\
            \phi &\sim \text{Unif}(0,1 - (m-1)\tilde\phi) \\
            \alpha_i,\beta,\gamma_1,\gamma_2 &\sim \text{N}(0,1)  \quad \text{for } i = 1,...,m \\
            \text{logit}(\pi) &\sim \text{N}(0,2^2). \\
    \end{aligned}
\end{equation}
This model is a Bayesian interpretation of the one used in \citet{held2005}, with an added underreporting mechanism. Thus, a discussion of prior specifications is needed. Note that if $\tilde\lambda_{i,t}$ is a column vector where $\tilde\lambda_{\cdot,t} = (\tilde\lambda_{1,t}, \tilde\lambda_{2,t},\tilde\lambda_{3,t},\tilde\lambda_{4,t})$, and similarly for $n_{\cdot,t}$, $\nu_{\cdot,t}$, and $Z_{\cdot,t-1}$, then $\tilde\lambda_{\cdot,t} = n_{\cdot,t} \odot \nu_{\cdot, t} + \Lambda Z_{\cdot,t-1}$, for $\Lambda_{ij} = \begin{cases}\phi & i = j \\
    \tilde\phi & i\neq j\end{cases}$. 
Should the largest eigenvalue of $\Lambda$ be less than $1$, the time series is stationary \citep{held2005}. The priors on $\phi$ and $\tilde\phi$ in \ref{france} are chosen to specifically induce that the largest eigenvalue of $\Lambda$ be less than $1$. These priors are found by noting that $\Lambda = \phi I_m + \tilde\phi(J_m - I_m) = (\phi-\tilde\phi)I_m + \tilde\phi J_m$, where $I_m$ is the $m\times m$ identity matrix and $J_m$ is the $m \times m$ matrix of ones, has eigenvalues $\phi + (m-1)\tilde\phi$ and $\phi - \tilde\phi$. Since $\phi$ and $\tilde\phi$ are greater than $0$, the largest eigenvalue is $\phi + (m-1)\tilde\phi$. Note that $\phi + (m-1)\tilde\phi < 1 \Rightarrow (m-1)\tilde\phi < 1-\phi < 1$ and thus $0 < \tilde\phi < 1/(m-1)$. Also, $\phi + (m-1)\tilde\phi < 1 \Rightarrow 0 <\phi < 1-(m-1)\tilde\phi$. These priors are uninformative on these ranges and the other priors are inspired by those of \ref{saarland} and \citet{slater2025}. Estimated and reweighted incidence, cases not attributable to past cases or age groups, within-group and out-of-group reproduction, and the reporting probability are visualized in \autoref{france_fig}, with the same overall trends as the univariate study. Resampling occurs only twice, at time steps 82 and 125 (see \autoref{meningoESS_fig}), indicating that the surrogate model is an apt proposal. The unknown case counts are pulled up, the endemic component is pulled down, the within-group reproduction is pulled up, the out-of-group reproduction is pulled down, and the reporting probability is pulled down. Our adjustments to the estimated true case counts are visualized in \autoref{franceCHANGE_fig}, where the change as a proportion of the observed cases is roughly 10\% for the lower quantile, 25-30\% for the median, and 20\% for the upper quantile. Our results greatly differ from \citet{held2005}, as they provide maximum likelihood estimates of $\phi = 0.12$ and $\tilde\phi = -0.0004$ whilst we provide median estimates of $\phi = 0.69455$ and $\tilde\phi = 0.05387$. These inconsistencies are attributed to their false assumption that $\pi = 1$, where we provide a posterior median estimate of $\pi = 0.36886$, leading to their severe underestimation of $\phi$ and $\tilde\phi$ \citep{slater2025}. As with the univariate study, we conclude that the surrogate model underestimates the incidence and prevalence of the disease.

\subsection{Interpretation of results}

In both case studies, the importance smoother adjusted the surrogate model by increasing estimates of the unknown cases $\{X_{i,t}\}$, decreasing the reporting probability $\pi$, increasing the reproductive rate $\phi$, and decreasing the endemic component $\nu$. The changes in $\{X_{i,t}\}$ happen for two intuitive reasons: (i) the normal-normal approximation allows proposed values of $\{X_{i,t}\}$ to be less than $\{Y_{i,t}\}$, and (ii) the skewness of the normal and Poisson distributions. As the domain of $\tilde p(Z_{i,t}|Z_{<t})$ is only restricted to $(0.1,\infty)$, proposals of $Z_{i,t}$ may map to an $X_{i,t}$ smaller than its associated $Y_{i,t}$. The probability of observing $Y_{i,t} > X_{i,t}$ is zero, and thus, these samples are assigned a weight of zero and are withheld from the computed expectation. In other words, certain ``small" proposals of $X_{i,t}$ will be excluded from the importance weights, causing estimates to increase after adjustments. Secondly, although the Gaussian approximation matches the first and second moments, it has symmetric left and right tails which do not capture the right-skew of the Poisson distribution. Accordingly, for large $Y_{i,t}$, too much probability is assigned to values of $X_{i,t}$ below the median whilst the upper tail is assigned too little probability. Consequently, the posterior distribution under the surrogate model is shifted toward smaller values of $X_{i,t}$, thereby resulting in lower median estimates. Provided the estimates of $X_{i,t}$ are too small, it must be the case that estimates for $\pi$ are too large, as the value of $Y_{i,t}$ is fixed. This reduction in $\pi$ explains the adjustments on $\nu$ and $\phi$, since more severe underreporting will directly cause the underestimation of $\phi$ and an overestimation of $\nu$ \citep{slater2025}.

\section{Discussion}\label{sec:discussion}

In this work, we introduced a computationally efficient modelling framework for fitting thinned count autoregressions using latent Gaussian approximations and sequential importance sampling. As shown in the simulation study, the approach is particularly helpful for small mean time series. The case studies reveal how our method is directly applicable to real-world, multivariate epidemic curve reconstruction. 

However, improvements can be made to the computation of the importance weights. The grid approximation has the na\"{\i}ve assumption that $\tilde p(Z_{i,t} | Z_{<t})$ is uniformly distributed on the set $A_{X_{i,t}}$, which is clearly not the case as seen in \autoref{LGMAP_fig}. Computational efficiency may be improved using other quadrature methods, such as Clenshaw-Curtis, which reduces complexity using the fast Fourier transform \citep{trefethen2008}. Bias is also a key concern, where although importance sampling is an unbiased approach, the normalization procedure introduces bias. In recent work, \citet{cardoso2022} introduces a method to reduce this bias without imposing significant additional computational burden. There are also multiple layers of numerical bias introduced via HMC, Monte Carlo approximations of importance weights, quadrature error, and empirical estimation of expected values, which may cause error in our adjustments. The surrogate model need not be fit using HMC, either. Other methods, including particle MCMC and INLA, may be adapted for this use case. Other potential continuous surrogates could be used in place of the normal-normal approximation to prohibit instances where $X_{i,t} < Y_{i,t}$ and reduce skewness misspecifications. Such a potential surrogate may provide additional closed-form results, reducing the need to approximate. 

Our proposed modelling framework provides a strategy for conducting inference with thinned count autoregressive models for applications in ecological population modelling and infectious disease surveillance. By leveraging the computational benefits of latent Gaussian surrogate models and correcting their associated misspecifications, our framework enables practitioners to incorporate flexible mechanisms that recover features of the underlying partially observed count process.

\section*{Acknowledgments}

We acknowledge the support of the Natural Sciences and Engineering Research Council of Canada (NSERC) (RGPIN-2025-05007). J. Fingold is supported by the Queen Elizabeth II Graduate Scholarship in Science and Technology (QEII-GSST).

\section*{Data availability statement}

This article uses data available in the R packages \textit{surveillance}
\citep{surveillance} and \textit{hhh4under\-reporting} \citep{bracher2021}.

\bibliography{references}

\newpage
\appendix
\counterwithin{figure}{section}
\renewcommand{\thefigure}{\thesection.\arabic{figure}}
\setcounter{figure}{0} 

\counterwithin{table}{section}
\renewcommand{\thetable}{\thesection.\arabic{table}}
\setcounter{table}{0} 

\section{Wed Appendix: Tables for Simulation Study}
\begin{table}[H]
\small
\centering
\begin{tabular}{cccrrrrrrrrr}
\hline
\multirow{2}{*}{$\nu/\phi/\pi$}
& \multirow{2}{*}{$\mathbbm{E}[X_t]$}
& \multirow{2}{*}{$\mathbbm{E}[Y_t]$}
& \multicolumn{3}{c}{Surrogate} 
& \multicolumn{3}{c}{Grid} 
& \multicolumn{3}{c}{Quadrature} \\
\cline{4-12}
& & & \multicolumn{1}{c}{Low} 
& \multicolumn{1}{c}{Med} 
& \multicolumn{1}{c}{High}
& \multicolumn{1}{c}{Low} 
& \multicolumn{1}{c}{Med} 
& \multicolumn{1}{c}{High}
& \multicolumn{1}{c}{Low} 
& \multicolumn{1}{c}{Med} 
& \multicolumn{1}{c}{High} \\
\hline
5/0.40/0.40 & 8.33 & 3.33 & 79.6 & 68.4 & 36.7 & 92.0 & 78.7 & 38.4 & \textbf{92.6} & \textbf{79.1} & \textbf{39.5}\\ 
5/0.50/0.50 & 10.00 & 5.00 & 73.1 & 71.8 & 48.0 & \textbf{88.2} & \textbf{79.2} & \textbf{51.5} & 88.0 & 75.8 & 51.1\\ 
5/0.60/0.53 & 12.50 & 6.67 & 73.9 & 63.2 & 47.8 & \textbf{84.7} & \textbf{81.8} & 52.8 & \textbf{84.7} & 81.1 & \textbf{53.5}\\ 
5/0.70/0.50 & 16.67 & 8.33 & 72.2 & 74.3 & 50.3 & 86.2 & 84.1 & 62.5 & \textbf{86.6} & \textbf{87.5} & \textbf{63.2}\\ 
5/0.80/0.40 & 25.00 & 10.00 & 72.8 & 50.7 & 34.0 & 73.2 & 56.2 & 39.0 & \textbf{76.0} & \textbf{58.5} & \textbf{41.0}\\ 
5/0.90/0.23 & 50.00 & 11.67 & 42.8 & 33.8 & 27.6 & 47.3 & 34.5 & \textbf{28.9} & \textbf{47.8} & \textbf{35.4} & 28.7\\ 
10/0.40/0.80 & 16.67 & 13.33 & 62.0 & 30.9 & 4.7 & \textbf{81.1} & \textbf{36.0} & \textbf{12.5} & 79.6 & 31.7 & 10.2\\ 
10/0.50/0.75 & 20.00 & 15.00 & 62.8 & 35.9 & 10.3 & \textbf{75.4} & \textbf{45.5} & \textbf{16.4} & 72.9 & 41.0 & 14.1\\ 
10/0.60/0.67 & 25.00 & 16.67 & 59.0 & 39.8 & 21.3 & \textbf{74.9} & \textbf{58.2} & \textbf{34.0} & 73.9 & 54.8 & 32.2\\ 
10/0.70/0.55 & 33.33 & 18.33 & 66.1 & 67.8 & 45.0 & \textbf{83.2} & \textbf{83.5} & \textbf{56.0} & 82.6 & 83.3 & 55.8\\ 
10/0.80/0.40 & 50.00 & 20.00 & 67.4 & 60.5 & \textbf{38.2} & 67.5 & 61.1 & \textcolor{red}{35.0} & \textbf{68.3} & \textbf{61.6} & \textcolor{red}{34.8}\\ 
10/0.90/0.22 & 100.00 & 21.67 & 34.7 & 30.5 & 20.5 & \textbf{39.1} & \textbf{31.1} & 21.4 & 39.0 & \textbf{31.1} & \textbf{21.5}\\ 
10/0.75/0.58 & 40.00 & 23.33 & 62.4 & 62.7 & 42.5 & \textbf{78.7} & \textbf{73.5} & \textbf{53.9} & 77.8 & 72.8 & 53.8\\ 
10/0.80/0.50 & 50.00 & 25.00 & 66.5 & 67.6 & 49.5 & \textbf{82.4} & 79.7 & 55.5 & \textbf{82.4} & \textbf{80.1} & \textbf{55.9}\\ 
10/0.85/0.40 & 66.67 & 26.67 & \textbf{70.9} & \textbf{62.9} & 39.0 & \textcolor{red}{70.4} & \textcolor{red}{57.3} & \textbf{39.2} & \textcolor{red}{70.6} & \textcolor{red}{57.8} & 39.0\\ 
10/0.75/0.71 & 40.00 & 28.33 & 54.9 & 32.6 & 14.0 & \textbf{66.2} & \textbf{48.0} & \textbf{26.7} & 65.4 & 47.0 & 26.6\\ 
10/0.80/0.60 & 50.00 & 30.00 & 51.5 & 46.8 & 31.3 & \textbf{68.7} & \textbf{61.8} & 43.5 & 68.0 & 61.6 & \textbf{43.7}\\ 
10/0.85/0.48 & 66.67 & 31.67 & 64.0 & 72.6 & 47.7 & 69.4 & 76.2 & \textbf{53.8} & \textbf{69.9} & \textbf{76.3} & \textbf{53.8}\\ 
10/0.75/0.83 & 40.00 & 33.33 & 61.8 & 37.5 & 10.5 & \textbf{71.7} & \textbf{47.7} & \textbf{18.2} & 70.7 & 46.1 & 17.6\\ 
10/0.80/0.70 & 50.00 & 35.00 & 56.3 & 37.1 & 20.2 & \textbf{71.4} & \textbf{54.3} & 34.9 & 71.1 & 53.8 & \textbf{35.0}\\ 
10/0.85/0.55 & 66.67 & 36.67 & 62.7 & 60.5 & 38.3 & 74.3 & \textbf{69.2} & \textbf{43.5} & \textbf{74.4} & 68.9 & 43.4\\ 
10/0.75/0.96 & 40.00 & 38.33 & 35.6 & \textbf{50.8} & 10.6 & \textbf{94.0} & \textcolor{red}{42.5} & 14.7 & 93.6 & \textcolor{red}{38.8} & \textbf{15.6}\\ 
10/0.80/0.80 & 50.00 & 40.00 & 60.7 & 37.6 & 16.7 & \textbf{70.4} & \textbf{48.5} & \textbf{24.7} & 69.2 & 47.5 & 24.5\\ 
10/0.85/0.62 & 66.67 & 41.67 & 55.2 & 50.6 & 38.8 & \textbf{70.4} & 59.8 & 47.6 & 70.2 & \textbf{60.0} & \textbf{48.0}\\ 
\hline
\text{Average} & & & 61.2 & 52.0 & 31.0 & \textbf{74.2} & \textbf{60.4} & \textbf{37.7} & 74.0 & 59.6 & 37.6 \\
\hline
\end{tabular}
\caption{Perfect match rates on 95\% credible intervals and medians for the surrogate model and for the importance sampling adjustments under both weight schemes. The match rate of the best-performing model is in bold for each of the lower, median, and upper quantiles for each parameter set, coloured in red if the approximate model is best. The surrogate, grid-adjusted, and quadrature-adjusted models had the best perfect match rates 4, 43, and 29 times, respectively. The grid scheme and the quadrature scheme both performed poorer than the surrogate model 4 times.}
\label{perfect_match_rates}
\end{table}

\begin{table}
\small
\centering
\begin{tabular}{cccrrrrrrrrr}
\hline
\multirow{2}{*}{$\nu/\phi/\pi$}
& \multirow{2}{*}{$\mathbbm{E}[X_t]$}
& \multirow{2}{*}{$\mathbbm{E}[Y_t]$}
& \multicolumn{3}{c}{Surrogate} 
& \multicolumn{3}{c}{Grid} 
& \multicolumn{3}{c}{Quadrature} \\
\cline{4-12}
& & & \multicolumn{1}{c}{Low} 
& \multicolumn{1}{c}{Med} 
& \multicolumn{1}{c}{High}
& \multicolumn{1}{c}{Low} 
& \multicolumn{1}{c}{Med} 
& \multicolumn{1}{c}{High}
& \multicolumn{1}{c}{Low} 
& \multicolumn{1}{c}{Med} 
& \multicolumn{1}{c}{High} \\
\hline
5/0.40/0.40 & 8.33 & 3.33 & 10.2 & 17.5 & 56.2 & 61.1 & 39.0 & 21.4 & \textbf{64.0} & \textbf{40.2} & \textbf{24.6}\\ 
5/0.50/0.50 & 10.00 & 5.00 & 13.4 & 15.4 & 36.0 & \textbf{55.7} & \textbf{32.6} & \textbf{15.9} & 55.1 & 21.7 & 13.7\\ 
5/0.60/0.53 & 12.50 & 6.67 & 13.0 & 19.9 & 39.9 & 41.1 & \textbf{53.8} & 21.0 & \textbf{41.2} & 52.6 & \textbf{23.2}\\ 
5/0.70/0.50 & 16.67 & 8.33 & 13.9 & 13.3 & 32.4 & 50.3 & 40.0 & 34.1 & \textbf{51.9} & \textbf{53.1} & \textbf{35.7}\\ 
5/0.80/0.40 & 25.00 & 10.00 & 13.6 & 26.9 & 54.2 & 1.3 & 15.0 & 26.1 & \textbf{11.4} & \textbf{20.1} & \textbf{28.7}\\ 
5/0.90/0.23 & 50.00 & 11.67 & 36.2 & 49.8 & 78.4 & 18.3 & 20.3 & 33.2 & \textbf{19.2} & \textbf{21.3} & \textbf{33.4}\\ 
10/0.40/0.80 & 16.67 & 13.33 & 19.0 & 38.3 & 94.8 & \textbf{50.3} & \textbf{8.1} & \textbf{14.1} & 46.4 & 1.3 & 10.5\\ 
10/0.50/0.75 & 20.00 & 15.00 & 18.7 & 35.0 & 75.7 & \textbf{34.3} & \textbf{20.5} & \textbf{22.1} & 27.6 & 12.8 & 19.7\\ 
10/0.60/0.67 & 25.00 & 16.67 & 20.8 & 32.7 & 55.2 & \textbf{39.6} & \textbf{33.3} & \textbf{25.7} & 37.3 & 28.7 & 23.3\\ 
10/0.70/0.55 & 33.33 & 18.33 & 16.9 & 18.0 & 34.7 & \textbf{50.6} & \textbf{54.2} & \textbf{30.3} & 49.1 & 53.7 & 30.2\\ 
10/0.80/0.40 & 50.00 & 20.00 & 16.3 & 20.7 & 42.5 & \textcolor{red}{-0.2} & 2.2 & \textbf{4.5} & \textbf{2.2} & \textbf{3.3} & \textbf{4.5}\\ 
10/0.90/0.22 & 100.00 & 21.67 & 46.5 & 56.6 & 96.7 & \textbf{21.9} & \textbf{23.2} & \textbf{28.4} & \textbf{21.9} & 23.1 & 28.3\\ 
10/0.75/0.58 & 40.00 & 23.33 & 19.2 & 19.3 & 35.8 & \textbf{43.3} & \textbf{30.5} & 19.9 & 41.0 & 28.4 & \textbf{20.1}\\ 
10/0.80/0.50 & 50.00 & 25.00 & 16.7 & 17.1 & 30.0 & \textbf{47.2} & 40.5 & 18.9 & \textbf{47.2} & \textbf{41.5} & \textbf{19.9}\\ 
10/0.85/0.40 & 66.67 & 26.67 & 14.6 & 19.3 & 40.4 & \textcolor{red}{-1.8} & \textcolor{red}{-11.6} & \textbf{6.0} & \textcolor{red}{-1.0} & \textcolor{red}{-10.4} & 5.7\\ 
10/0.75/0.71 & 40.00 & 28.33 & 23.1 & 36.3 & 64.5 & \textbf{23.2} & \textbf{26.0} & 26.3 & 22.0 & 24.7 & \textbf{26.4}\\ 
10/0.80/0.60 & 50.00 & 30.00 & 25.4 & 28.7 & 41.8 & \textbf{38.3} & \textbf{31.2} & 24.4 & 37.1 & 31.0 & \textbf{24.5}\\ 
\textbf{}10/0.85/0.48 & 66.67 & 31.67 & 18.5 & 14.2 & 31.7 & 16.0 & 16.2 & 14.2 & \textbf{17.4} & \textbf{16.5} & 14.0\\ 
10/0.75/0.83 & 40.00 & 33.33 & 19.6 & 33.8 & 70.8 & \textbf{27.2} & \textbf{16.8} & \textbf{26.0} & 24.7 & 15.2 & 25.9\\ 
10/0.80/0.70 & 50.00 & 35.00 & 22.3 & 33.0 & 53.8 & \textbf{35.5} & \textbf{27.7} & 26.3 & 34.7 & 26.9 & \textbf{26.5}\\ 
10/0.85/0.55 & 66.67 & 36.67 & 19.4 & 21.1 & 39.5 & 33.0 & \textbf{25.7} & 13.9 & \textbf{33.1} & 25.2 & \textbf{14.0}\\ 
10/0.75/0.96 & 40.00 & 38.33 & 32.1 & 25.8 & 85.7 & \textbf{90.6} & \textcolor{red}{-20.7} & 27.0 & 90.0 & \textcolor{red}{-27.0} & \textbf{29.7}\\ 
10/0.80/0.80 & 50.00 & 40.00 & 20.2 & 32.7 & 56.2 & \textbf{26.5} & \textbf{19.9} & 22.6 & 23.6 & 18.4 & \textbf{22.7}\\ 
10/0.85/0.62 & 66.67 & 41.67 & 23.3 & 25.9 & 38.4 & \textbf{35.2} & 20.8 & 21.4 & 34.6 & \textbf{21.1} & \textbf{22.0}\\ 
\hline
\text{Average} & & & 20.5 & 27.1 & 53.6 & \textbf{34.9} & \textbf{23.5} & 21.8 & 34.7 & 22.6 & \textbf{22.0} \\
\hline
\end{tabular}
\caption{Distance to perfect match rates on 95\% credible intervals and medians for the surrogate model and percentage in that distance lost under importance sampling adjustments models using both weight schemes. The match rate of the best-performing model is in bold for each of the lower, median, and upper quantiles for each parameter set. Instances where the adjustments add distance, and hence the importance weights decrease accuracy, are coloured in red. The grid weights perform best 38 times and the quadrature weights perform best 33 times. Distance is added on 4 occasions for the grid scheme and on 3 occasions for the quadrature scheme.}
\label{dist_match_rates}
\end{table}

\newpage
\section{Web Appendix: Figures for Case Studies}
\begin{figure}[H]
    \centering
    \includegraphics[width=\linewidth]{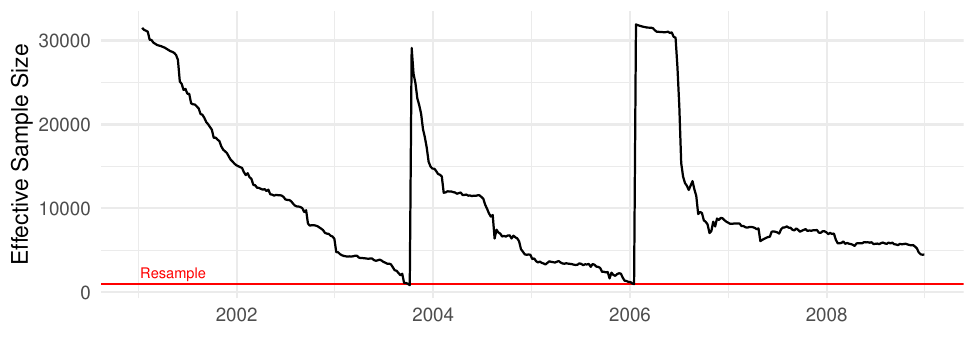}
    \caption{Effective sample size over time for univariate case study. The red horizontal line corresponds to the resampling threshold. Resampling occurred at $t=144, 263$, corresponding to 2003-10-06 and 2006-01-16.}
    \label{saarlandESS_fig}
\end{figure}

\begin{figure}[H]
    \centering
    \begin{subfigure}{\textwidth}
        \centering
        \includegraphics[width=\textwidth]{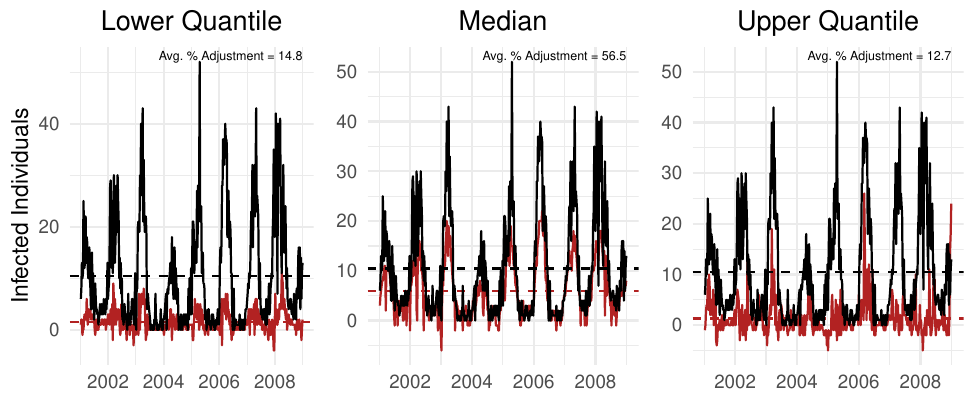}
    \end{subfigure}
    \begin{subfigure}{\textwidth}
        \centering
        \includegraphics[width=\textwidth]{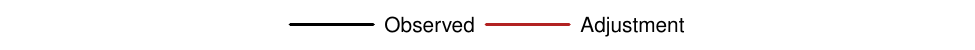}
    \end{subfigure}
    \caption{Importance sampling adjustments on estimates of X for each of the lower, median, and upper quantiles. The black dotted line corresponds to the average number of observed cases, and the red line corresponds to the average adjustment from the estimates found from \ref{saarland} and the importance sampled estimates. The Avg. \% Adjustment refers to the average adjustment in $X$ as a proportion of $Y$.}
    \label{saarlandCHANGE_fig}
\end{figure}

\begin{figure}[H]
    \centering
    \includegraphics[width=\linewidth]{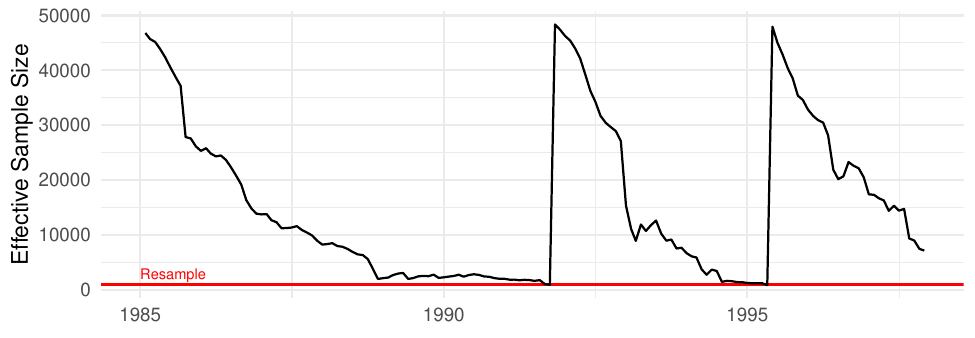}
    \caption{Effective sample size over time for multivariate case study. The red horizontal line corresponds to the resampling threshold. Resampling occurred at $t=82, 125$, corresponding to 1991-10-01 and 1995-05-01.}
    \label{meningoESS_fig}
\end{figure}

\begin{figure}[H]
    \centering

    \begin{subfigure}{\textwidth}
        \centering
        \includegraphics[width=\textwidth]{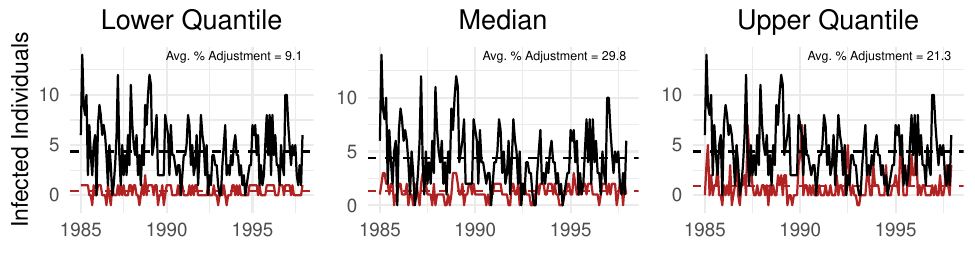}
        \caption{Age $<$ 1}
    \end{subfigure}

    \begin{subfigure}{\textwidth}
        \centering
        \includegraphics[width=\textwidth]{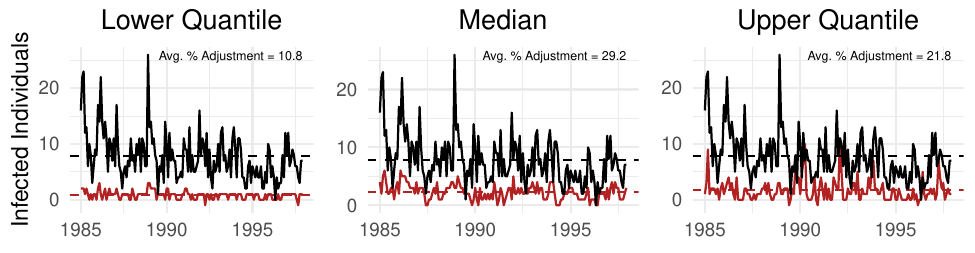}
        \caption{Ages 1-5}
    \end{subfigure}

    \begin{subfigure}{\textwidth}
        \centering
        \includegraphics[width=\textwidth]{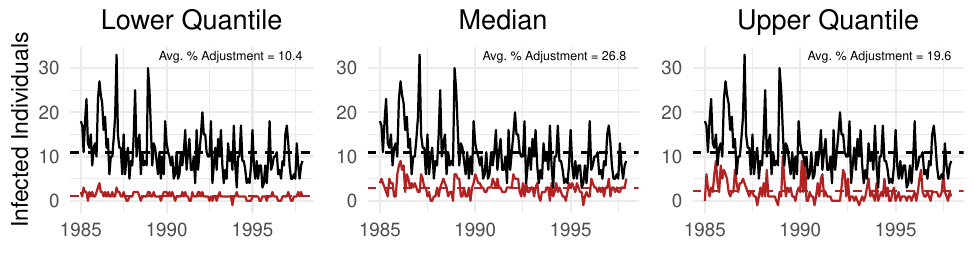}
        \caption{Ages 5-20}
    \end{subfigure}

    \begin{subfigure}{\textwidth}
        \centering
        \includegraphics[width=\textwidth]{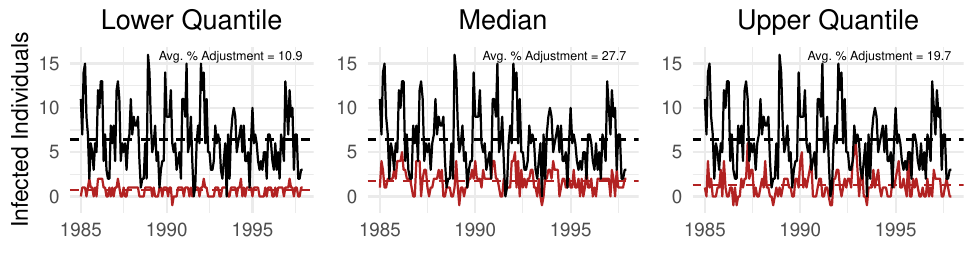}
        \caption{Age $>$ 20}
    \end{subfigure}
    \begin{subfigure}{\textwidth}
        \centering
        \includegraphics[width=\textwidth]{LEGEND_CHANGE.pdf}
    \end{subfigure}

    \caption{Importance sampling adjustments on estimates of X for each of the lower, median, and upper quantiles from \ref{france} for each age group. The Avg. \% Adjustment refers to the average adjustment in $X$ as a proportion of $Y$.}
    \label{franceCHANGE_fig}
\end{figure}

\end{document}